\documentclass[myepj-spec,pdftex]{mySvjour}
\usepackage{graphicx}
\usepackage{hyperref}
\usepackage{color}
\usepackage{xspace}
\usepackage[square,sort&compress,numbers]{natbib}   
\usepackage{verbatim}
\usepackage{amsmath,amssymb,amsfonts} 
\usepackage{tabularx}
\usepackage{multicol}
\usepackage{footnote}
\usepackage{listings}
\usepackage{graphicx} 
\usepackage{xcolor}
\usepackage{tikz}
\usetikzlibrary{positioning,arrows.meta}
\tikzset{
  box/.style   = { rectangle, rounded corners, draw},
  valve/.style = {-{Triangle[fill=white,width=1em,
      length=1em]Triangle[fill=white,width=1em,length=1em,reversed]}},
}
\tikzstyle{arrow} = [->,>=stealth]
\usepackage[switch, modulo]{lineno}

\usepackage{lmodern,bm}                
\usepackage[T1]{sansmath} 
\SetMathAlphabet{\mathsfbf}{sans}{\sansmathencoding}{\sfdefault}{bx}{sl}
\usepackage{etoolbox}
\AtBeginEnvironment{sansmath}{}{}{}

\usepackage{lipsum}

\newcommand{\Supax}{\textsc{Supax\,}}

\newcommand{\scanningSpeed}{0.004\,\text{MHz/h  to\,}0.11\,\text{MHz/h}}
\newcommand{\scanningSpeedSC}{0.05\,\text{MHz/h  to\,} 0.65\,\text{MHz/h}}
\newcommand{\TuningSlopeNbN}{$-27.45\,$kHz/mb}
\newcommand{\AxionMassMin}{8\,$\mu$eV\,}
\newcommand{\AxionMassMax}{30\,$\mu$eV}

\tikzset{
  box/.style   = { rectangle, rounded corners, draw},
  valve/.style = {-{Triangle[fill=white,width=1em,
      length=1em]Triangle[fill=white,width=1em,length=1em,reversed]}},
}
\tikzstyle{arrow} = [->,>=stealth]

\newcommand{\fres}{8.471~GHz }
\newcommand{\scanRange}{300~kHz }
\newcommand{\Qnbn}{\ensuremath{3.46\cdot 10^5} }
\newcommand{\Qcu}{\ensuremath{4.6\cdot 10^4} }

\definecolor{darkblue1}{rgb}{0,0,.2}
\definecolor{darkblue}{rgb}{0,0,.2}
\definecolor{darkred}{rgb}{0.5,0,0}
\pagecolor{white} 
\color{black}     
\hypersetup{breaklinks=true, 
	colorlinks=true, 
	linkcolor=darkblue1, 
	menucolor=darkblue1, 
	urlcolor=darkblue1,
	citecolor=darkblue1,
	pdftitle={},
	pdfauthor={},
	pdfsubject={},
	pdfkeywords={},
	pdfproducer={}
}
\usepackage{siunitx}

\bibstyle{plain}

\begin{document}
	
	{%
		\begin{@twocolumnfalse}
			
			\begin{flushright}
				\normalsize
			\end{flushright}
			
			\vspace{-2cm}
			\title{\Large\boldmath Conceptual Design Report of the SUPAX Experiment}

\author{Dhruv Chouhan$^{1}$, Tim Schneemann$^{2}$, Kristof Schmieden$^{1}$, Matthias Schott$^{1}$}
\institute{
\inst{1} Rheinische Friedrich-Wilhelms-University, Bonn, Germany \\ \inst{2} PRISMA+ Cluster of Excellence, Institute of Physics, Johannes Gutenberg University, Mainz, Germany 
}

			\abstract{The SUPerconduction AXion search experiment (\Supax) is a haloscope designed to probe axion-like particles (ALPs) as candidates for dark matter and solutions to the strong CP problem. ALPs are predicted to couple to photons, allowing their detection through resonant conversion in electromagnetic cavities placed within strong magnetic fields. \Supax employs a 12 T magnetic field and tunable superconducting cavities with resonance frequencies ranging from 2\,GHz to 7.2\,GHz, enabling the exploration of axion masses between \AxionMassMin and \AxionMassMax. The tuning mechanism, based on piezo motors and gas-pressure regulation, allows for simultaneous scanning of up to three frequencies, significantly improving search efficiency. This paper presents the technical design of the \Supax experiment, preliminary R\&D efforts, and results from prototype experiments.  In particular, we exclude dark photons with masses around $35\,\mu$eV with a kinetic mixing parameter $\chi > 5\cdot 10^{-14}$, i.e. a region of parameter space which has not been previously explored.}	
	\maketitle
	\end{@twocolumnfalse}
}

\tableofcontents


\section{Introduction}

Axion-like particles (ALPs) have emerged as a compelling extension to the Standard Model, offering potential solutions to fundamental problems in particle physics and cosmology. Initially proposed to solve the strong CP problem in Quantum Chromodynamics (QCD) \cite{PhysRevLett.38.1440,PhysRevLett.40.223,PhysRevLett.40.279}, axions and their generalizations, ALPs, also serve as viable candidates for dark matter. Their existence is motivated by various high-energy theoretical frameworks, and their interactions with photons provide an accessible experimental avenue for their detection. A detailed review can be found in \cite{Graham:2015ouw, Adams:2022pbo}.

ALPs generically couple to photons through an interaction term of the form $\mathcal{L}_{int} = -\frac{g_{a\gamma\gamma}}{4}aF\tilde{F} = -g_{_{a\gamma\gamma}}a\mathbf{E}\cdot\mathbf{B}$, where $a$ represents the ALP field, $g_{a\gamma\gamma}$ is the model dependent axion-photon coupling constant, and $F$ and $\tilde{F}$ denote the electromagnetic field tensor and its dual and $\mathbf{E}\cdot\mathbf{B}$ is the scalar product between the magnetic end electric fields. Two benchmark models, known as KSVZ \cite{Shifman:1979if,Kim:1979if} and DFSZ \cite{Dine:1981rt} yield coupling values between -0.97 and 0.36, respectively. This coupling underlies a wide range of experimental efforts to detect ALPs across different mass ranges and coupling strengths. The parameter space for the QCD axion remains vast, spanning many orders of magnitude in both mass and coupling strength, and similar considerations apply to ALP dark matter candidates.

Several experimental approaches have been developed to probe ALPs. Model-independent techniques such as "light-through-wall" experiments and helioscopes exploit ALP-photon conversion in controlled laboratory environments \cite{OSQAR:2015qdv, Ehret:2010mh}. A more targeted approach, assuming that ALPs constitute dark matter, involves haloscope experiments. These traditional searches employ resonant electromagnetic cavities placed in strong magnetic fields. If an ALP mass matches the cavity's resonance frequency, ALPs can convert into detectable photons. However, a key challenge is that the resonance frequency must be tunable to explore a wide range of possible ALP masses \cite{ADMX:2010ubl, Brubaker:2018ebj, QUAX:2024fut, Ahyoune:2024klt}. In recent years, alternative broad-band detection concepts like MadMax \cite{MADMAXinterestGroup:2017koy} have been proposed to circumvent the tuning limitation. While these approaches hold promise, they present significant technical challenges that remain to be fully addressed. 

The SUPerconduction AXion search experiment (Supax) follows the traditional haloscope methodology, aiming to explore axion masses between \AxionMassMin{} and \AxionMassMax{} with axion-photon couplings down to $g_{a\gamma\gamma}\lessapprox 2\cdot 10^{-15}$\,GeV$^{-1}$. The experiment leverages a 12\,T magnetic field and tunable superconducting cavities operating at resonance frequencies between 2\,GHz and 7.2\, GHz. Frequency tuning is achieved via piezo motors and gas-pressure regulation, allowing simultaneous scanning of multiple frequencies.

The structure of this paper is as follows. Section 2 presents the technical design of the \Supax experiment, which builds on extensive feasibility studies conducted over the past years. Section 3 details preliminary research and development efforts along with prototype experiments. Finally, Section 4 discusses the results obtained from the prototype experiment and their implications for future developments in ALP searches.

\section{Technical Design Specifications}

This section details the technical design of the final \Supax experiment. It will comprise a closed cycle cryostat operating at a base temperature of 10\,mK.  The experimental volume will be immersed in a 12\,T solenoidal magnetic field. A field cancellation is foreseen above the magnet housing the first stage of the readout electronics.

\subsection{Magnet System and Cryostat}

The \Supax experiment employs a state-of-the-art commercially available dilution refrigerator system, designed to provide ultra-low temperatures required for axion-like particle searches. The system is optimized for high cooling power, stability, and integration with the 12\,T magnet, ensuring an efficient experimental setup. The dilution refrigerator has a compact design when closed, with a height of 2 meters and a width of 1.7 meters. When fully opened, the system extends to a height of 3.2 meters with a diameter of 1.5 meters, allowing for convenient access to internal components during assembly and maintenance.

The system operates with three cooling stages, progressively lowering the temperature to the mK regime. The third cooling stage, which is critical for the experiment, provides cooling at 20\,mK and supports a heat load of $Q_{20mK} > 17 \mu W$. At 100\,mK the cooling power increases to $Q_{100mK} > 500 \,\mu W$.

The lowest cooling stage, where the magnet as well as the electromagnetic cavity is housed, features a cylindrical volume with a height of 550\,mm and a diameter of approximately 400\,mm. This stage ensures stable and uniform cooling for the experiment’s sensitive components, maintaining ultra-low temperatures necessary for the detection of weak axion-photon signals. A 12\,T superconducting solenoid magnet is incorporated directly within the dilution refrigerator. The magnet has a 90\,mm bore and provides a field homogeneity of 10\% within a 200\,mm region centered around the magnet. 

For efficient data acquisition and control, several electronic connections are integrated into the system, enabling signal transmission from the magnet volume to the outer cryostat cooling stages. Two dedicated lines allow for the operation of Josephson Parameteric Amplifiers (JPAs) as first  amplification stage, critical for detecting weak electromagnetic signals. Moreover, two electrical as well as one thermally shielded rod connection across the three cooling stages are foreseen. In addition, multiple electrical feedthroughs enable real-time temperature monitoring and fine thermal control across different cooling stages will be integrated. Superconducting coaxial cables are routed through the mixing flange, providing low-noise paths for extracting photon signals generated within the cavity.

\subsection{Cavity Design and Frequency Tuning}

\begin{figure}[tb]
\centering
\begin{minipage}{7.2cm}
  \centering
    \includegraphics[width=0.99\linewidth]{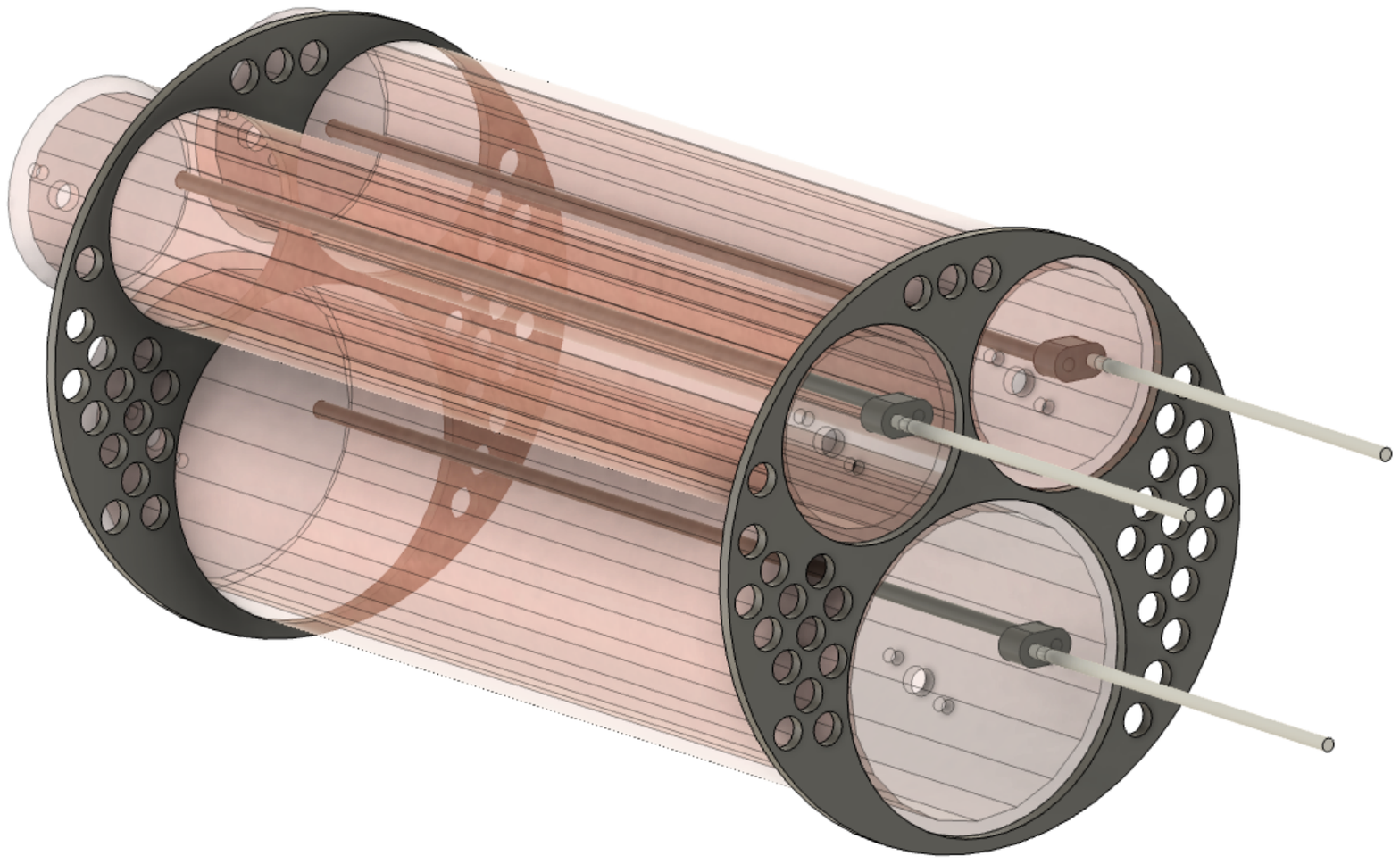}
    \caption{CAD Drawing of the three-cavity design of the \Supax experiment including holding structure and tuning mechanism.\vspace{0.6cm}}
    \label{fig:CavityDesign}
\end{minipage}%
\hspace{0.2cm}
\begin{minipage}{9.2cm}
  \centering
    \includegraphics[width=0.99\linewidth]{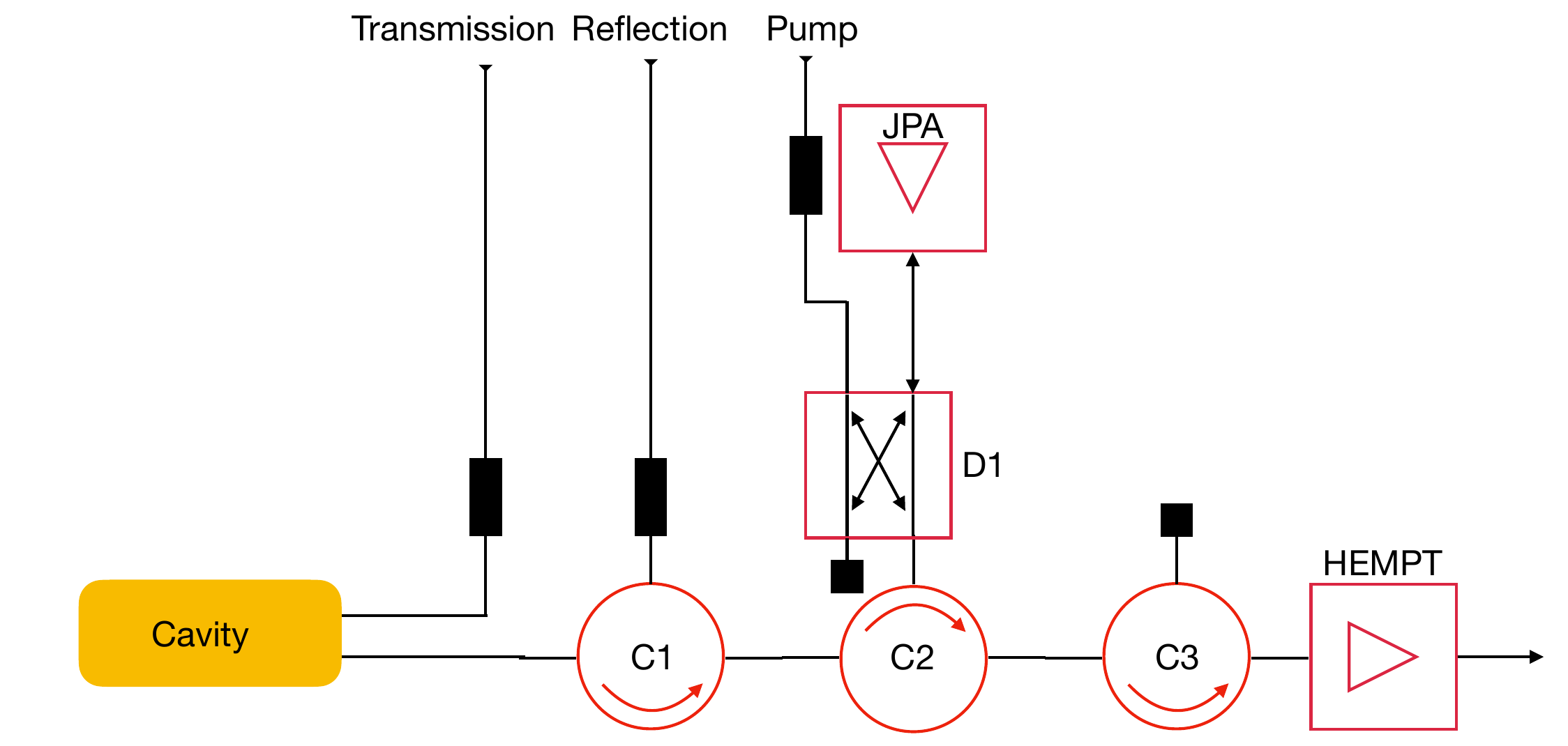}
    \caption{Schematic view of the cryogenic RF layout. The receiver signal path is shown in red. The calibration signal path is shown in black. }
    \label{fig:cryoRFLayout}
\end{minipage}
\end{figure}

The cavity design of the \Supax experiment is schematically illustrated in Figure \ref{fig:CavityDesign}. The cavity is housed within the available magnetic field volume, which is cylindrical with a 90 mm diameter and a height of 200 mm. The \Supax cavity-body is positioned within this volume to maximize sensitivity.

Three cylindrical cavities are combined within a common holding structure. These cavities are precisely machined using a CNC process, starting from both ends of the copper body. Each cavity has a base shaped as a tetracontagon (a 40-sided polygon) and has a different diameter between 3.7\,cm and 1.7\,cm to achieve different resonance frequencies. The length of the cavities is 18 cm. A dielectric rod with 5\,mm diameter is placed in each cavity off center, connected to a movement system allowing to pivot the rod towards the enter of the resonant volume achieving a change in frequency. The tuning range is up to 1\,GHz, depending on the diameter of the cavity.  

The motion of tuning element is realized through a steering rod, which connects the internal rod to an external piezo motor system. The steering rod is guided through the cavity structure and cryostat, bypassing the high-field central region of the solenoid. It reaches out toward the outer parts of the magnetic field, where the magnetic field strength is sufficiently low to allow safe and interference-free operation of the piezoelectric actuators.
This setup allows for highly precise, cryo-compatible positioning of the rod with micron-level resolution. The fixed endcaps on the top and bottom of the copper block are coated with ReBCO tapes, as are the 40 rectangular walls formed by sides of the tetracontagon, ensuring superconducting performance across the entire inner surface. These tapes significantly reduce surface resistance and are key to reaching a target quality factor of $Q_0 \approx 4.5 \times 10^5$. In the initial experimental phase, the cavities will be machined from bulk copper, with an expected quality factor around $Q_0 \approx 60{,}000$. The integration of ReBCO coatings is planned as a future upgrade to enhance performance.

For fine-tuning, $^3$He gas pressure within the cryostat is regulated between 0 to 1 mbar at temperature ranges where the He remains gaseous. Adjusting the gas pressure alters the dielectric constant inside the cavities, thereby refining the resonance conditions and providing an additional fine-tuning capability for resonance frequency adjustments.

The $TM_{010}$ mode exhibits the largest sensitivity to the axion field and is chosen as readout mode.  
The tripple-cavity setup offers several advantages. The ability to scan multiple resonance frequencies simultaneously while optimally using the available magnetized volume significantly improves the search efficiency for axions. Additionally, the combination of a simple mechanical structure for coarse tuning and a gas-based fine-tuning mechanism ensures both flexibility and precision in frequency adjustments. These design choices are based on extensive preliminary studies, as discussed in Section \ref{sec:freq-tuning}. Each cavity is equipped with two ports, one critically coupled and one very weakly coupled to the target cavity mode. In all designs the length of the strongly coupled antenna is adjusted via a piezo driven actuator, which is able to move the antenna port in a range of $\pm 0.5$\,mm. 

To scan the mass range envisioned between 10\,$\mu$eV to 20\,$\mu$eV three tripple-cavities will be used.  

\subsection{Readout}

Each cavity is equipped with its own dedicated readout chain. Consequently, the readout system described below is implemented two times in parallel. The cryogenic RF layout is illustrated schematically in Fig. \ref{fig:cryoRFLayout}. 
The strongly coupled port of each cavity is connected to a Josephson Parametric Amplifier (JPA) via two circulators. The JPA provides a gain of 20\,dB. One circulator isolates the JPA from incoming signals reflected off the cavity, while the other separates the input and output signals of the JPA. A directional coupler is used to inject the JPA pump tone into the input line without interference. 

The output of the JPA is routed through a third circulator to a HEMT amplifier, located at the 4\,K stage of the cryostat. From there, the signal continues to the room-temperature electronics. To minimize thermal noise, the pump tone for the JPA is attenuated with 20\,dB attenuators at both the 4\,K and base temperature stages of the cryostat. All signal lines connecting the cavity, JPA, and HEMT input are implemented using superconducting cables to minimize losses.

The weakly coupled port of the cavity is connected to the Vector Network Analyzer (VNA) at room temperature via an isolator and a 20 dB attenuator at the 4\,K stage, enabling transmission measurements through the cavity. 
For reflection measurements at the strongly coupled port, the third port of circulator C1 is connected via a 20\,dB attenuator at base temperature, and an additional attenuator at 4\,K to the VNA at room temperature. The room-temperature electronics include a commercial pump-tone generator for the JPA, a Vector Network Analyzer (VNA), and a real-time spectrum analyzer with an integrated low-noise preamplifier.

\section{Expected Sensitivities}

The expected sensitivity of any cavity based axion search depends on the expected signal power $P_{sig}$ and the systems noise power $P_{noise}$. The power of the axion signal generated in the cavity can be calculated as follows

\begin{equation}
P_{\text{sig}} = \left( \frac{g_{a\gamma\gamma} \alpha_{\text{em}}}{\pi} \right)^2 \frac{(\hbar c)^3 \rho_a}{\Lambda^4} \frac{B^2 \beta}{(1 + \beta)^2} Q_0 V C_{010} \frac{2\pi f_0}{\mu_0},
\end{equation}
when the axion mass is close to resonant frequency of the cavity. The coupling of the antenna to the cavity mode is labeled $\beta$, the cavity's volume $V$ and the overlap of the cavity mode $T_{mnl}$ with the axion field $C_{mnl}$. $B$ is the magnetic field strength.
Most notably it is proportional the the Volume $V$ and the quality-factor $Q_0$  of the cavity as well as the resonance frequency $f_0$. 
The signal power also depends on the local axion density $\rho_a$ and the axion photon coupling $g_{a\gamma\gamma}$. 
For the following considerations the coupling values are taken from the KSVZ model. The expected signal power ranges therefore between $6\cdot 10^{-23}\,W$ and $1.7\cdot 10^{-22}\,W$ for the min. and max. cavity volumes of a superconducting cavity and a factor ten lower for a normal conducting copper cavity. 

The noise power from the cavity is calculated as $P_{noise} = k_B T_{sys} / \sqrt{\tau / \Delta v_a}$, where $\tau$ is the integration time and $\Delta v_a = m_a <v> / h$ is the bandwidth of the readout taken to be the line-width of the axion signal. The system noise temperature is calculated as

\begin{equation}
k_B T_{\text{sys}} = h \nu \left( \frac{1}{e^{h \nu / k_B T_{\text{amb}}} - 1} + \frac{1}{2} + N_A \right),
\end{equation}
where $T_{\text{amb}}$ is the ambient temperature of the setup and $N_A \ge 0.5$ is the added noise of the amplifier, where $N_A = 0.5$ corresponds to the standard quantum limit for sufficiently low temperatures, which will be reached with the proposed readout scheme. 
The noise power mainly depends on the system temperature. Consequently, at sufficiently low temperatures, quantum fluctuations become the dominant source of noise in the system. 
Minimizing this quantum noise is essential for enhancing the sensitivity of axion detection experiments. The future \Supax experiment will run at a temperature of 10\,mK. The scanning speed is adjusted so that the noise power is about one order of magnitude lower than the expected signal power. The SNR ratio is given as 
\begin{equation}
    \text{SNR} = \frac{P_{sig}}{k_BT_{sys}}\sqrt{\frac{\tau}{\Delta\nu_a}},
\end{equation}
where $\tau$ is the integration time, and $\Delta\nu_a = m_a\left<v^2\right>/h$ with $\left<v^2\right> \approx 10^{-3}c$. As shown in \cite{Rapidis:2018xcz} the scanning speed of the setup can be calculated as follows:
\begin{equation}
    \frac{d\nu}{dt} \approx \frac{4}{5} \frac{Q_L Q_a}{\text{SNR}^2}\left( g_\gamma^2 \frac{\alpha^2}{\pi^2}\frac{\hbar^3c^3\rho_a}{\Lambda^4}\right)^2
    \times\left( \frac{1}{\hbar\mu_0}\frac{\beta}{1+\beta}B_0^2VC_{mnl}\frac{1}{N_{sys}} \right)^2
\end{equation}
where $Q_L$ and $Q_a = 10^6$ are the loaded quality factor of the cavity and the effective quality factor of the axion signal, SNR is the signal to noise ratio of the setup. Here, $g_\gamma$ is a model-dependent dimensionless coupling that is related to the physical coupling $g_{a\gamma\gamma}$ appearing in the axion-photon lagrangian by $g_{a\gamma\gamma} = (g_\gamma\alpha/\pi\Lambda^2)m_a$, where $\alpha$ is the fine-structure constant, $\Lambda = 77.6\,\text{MeV}$ encodes the dependence of the axion mass on hadronic physics, $m_a$ is the axion mass and $\rho_a = 0.45\,\text{GeV/cm}^3$ is the local axion dark matter density.

Following the analysis used in the data-analysis approach for our preliminary studies (See Section \ref{sec:darkPhotonsAnalysis}) a null measurement can be converted into 95\% CL upper limit on the axion coupling.
With the goal to reach the QCD axion band in sensitivity, i.e. $g_{a\gamma\gamma}< g_{\text{KSVZ}}$, a scanning rate in frequencies around $dv/dt = \scanningSpeed{}$ and $dv/dt = \scanningSpeedSC{}$, for normal conducting ($Q_L = 40k$) and a superconducting cavities ($Q_L = 400k$) is reached, respectively.

\begin{figure}
    \centering
    \includegraphics[width=0.75\linewidth]{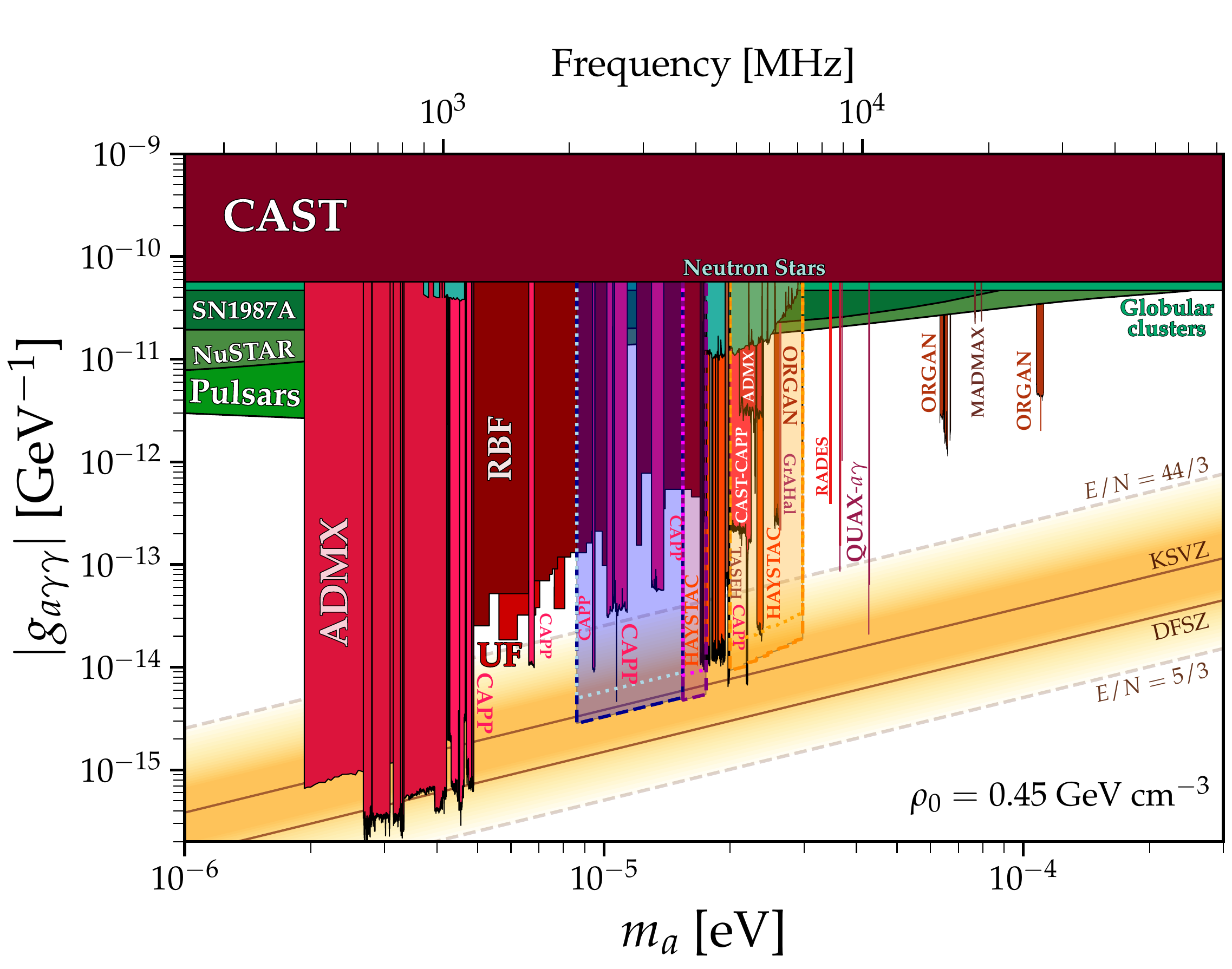}
    \caption{Projected sensitivity of the \Supax experiment for a measurement time of one year is indicated in the shaded areas for low (blue and purple) and high frequency (orange) frequency ranges using normal conducting copper cavities (dotted line) and superconducting cavities (dashed line). The scan of the purple area would take an additional half year to close the gap to the HAYSTAC measurement.}
    \label{fig:supax_II_sensitivity}
\end{figure}

Based on our experiences during the prototype constructions, we estimate a time-requirement of 20\,s to adjust the cavities for a new frequency, which also has to be taken into account for the final effective scanning speed. 
The magnet installations as well as the final setup of the cavity are expected to be finalized in spring 2026 with a subsequent start of data-taking from summer 2026 onwards. The data-taking will last for 80 weeks between 2026 and 2029. Within one year of operation a useful data taking time of 40 weeks is envisioned. The expected sensitivities of \Supax after the data-taking period are shown in Figure \ref{fig:supax_II_sensitivity}, where the achieved SNR has been converted into an upper limit on the axion coupling in the absence of an observed signal. 

\section{Prototype Experiment} \label{sec:prototype}
To gain initial experience with a haloscope-style experiment and to test novel cavity designs and superconducting materials for the \Supax experiment, a prototype setup has been developed and commissioned at the University of Mainz. The system operates using a liquid helium (LHe) flow cryostat, achieving a base temperature of 1.5\,K, and features a readout chain based on a HEMT amplifier.

Further details of this prototype setup are provided in the following section.

\subsection{Experimental Setup}

\begin{figure}[tb]
\centering
\begin{minipage}{6.2cm}
  \centering
    \includegraphics[width=0.99\linewidth]{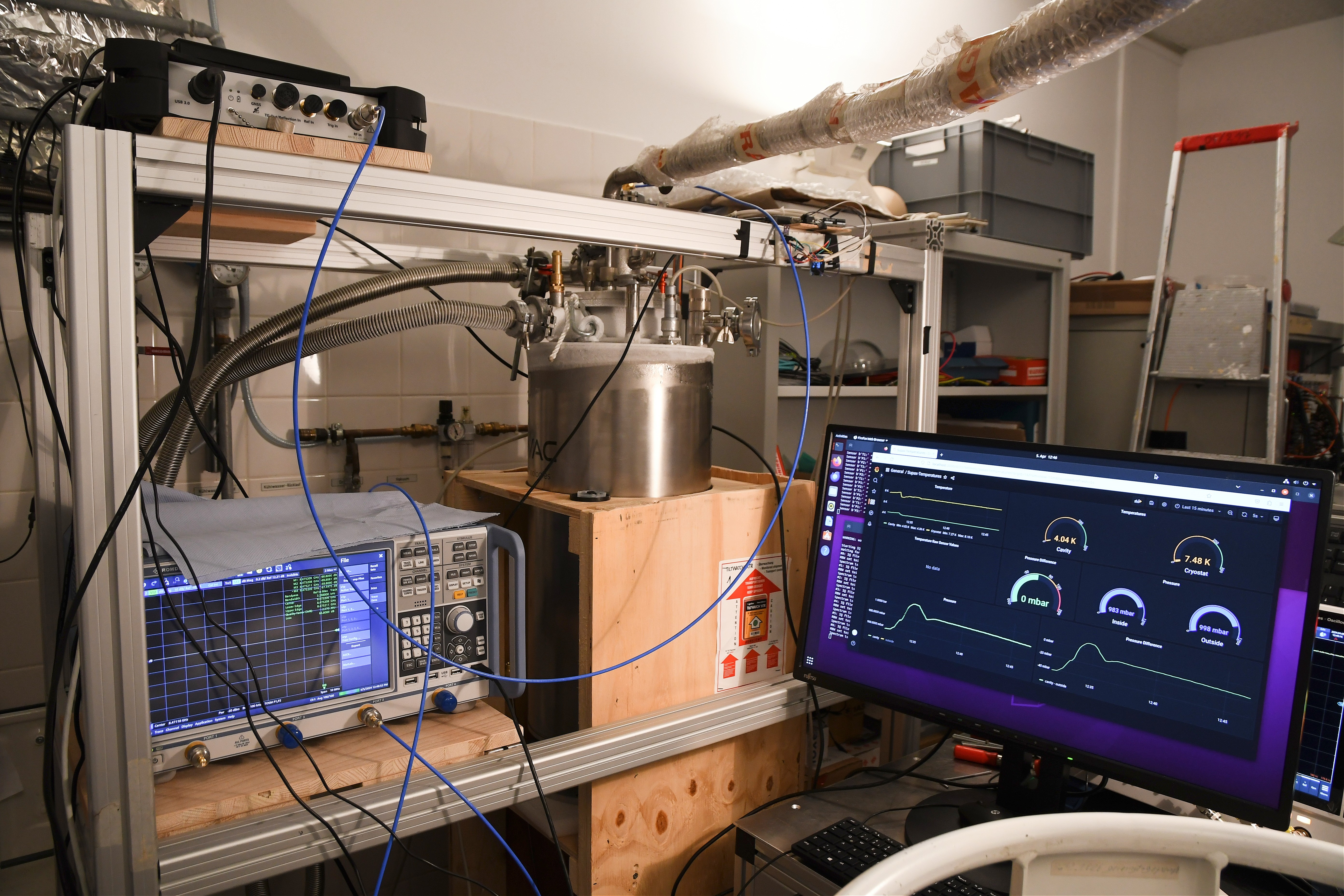}
    \caption{Picture of the \Supax experiment during initial tests. In The Background the cryostat is visible, to it's left a VNA and to its right a monitoeing display.}
    \label{fig:setup}

\end{minipage}%
\hspace{0.2cm}
\begin{minipage}{10.3cm}
  \centering
    \includegraphics[width=0.99\linewidth]{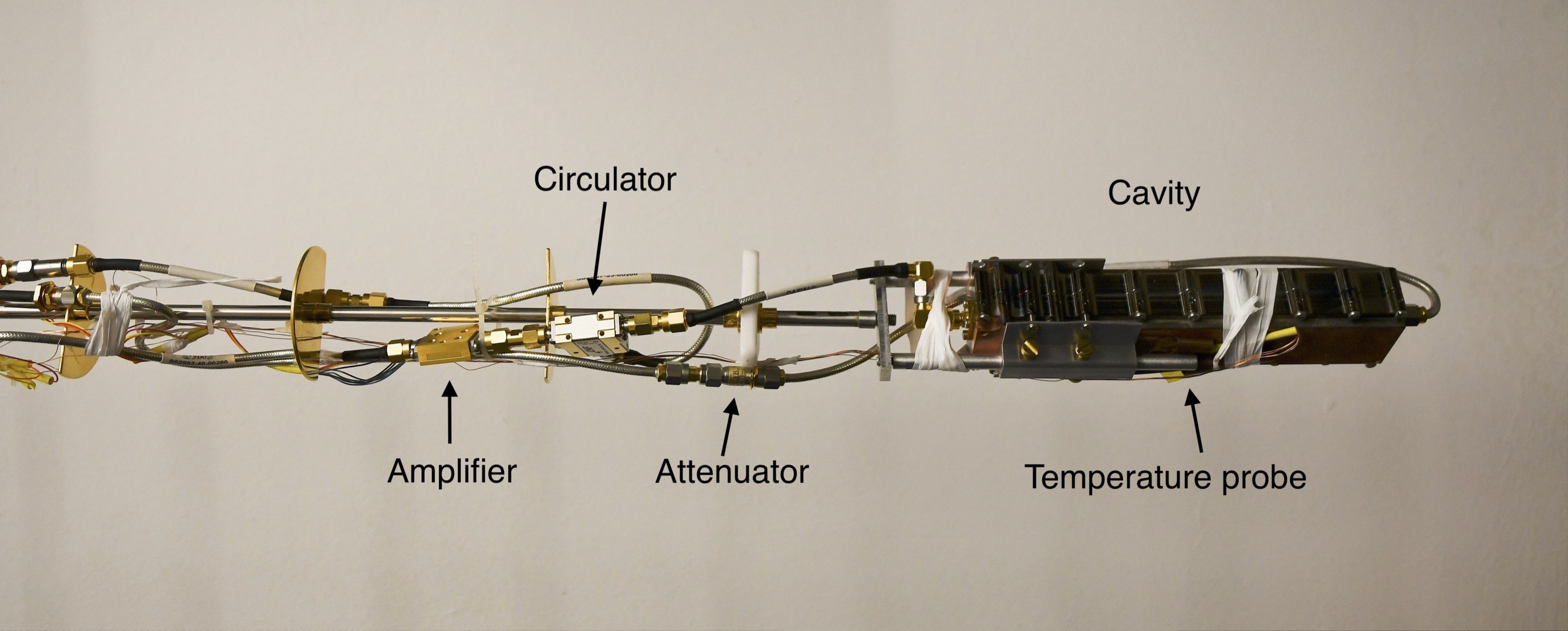}
    \caption{Annotated image of the cryostat inset with installed cavity and electronics. Not visible is the heat exchanger where the amplifier is attached to during operation.\vspace{0.35cm}}
    \label{fig:PictureInset}
\end{minipage}
\end{figure}

The core of the prototype experiment is a RF-cavity cooled to 2\,K in a dedicated cryogenic setup (Fig. \ref{fig:setup}).  The cavity is placed in a liquid helium flow cryostat. The lower part of the cryostat itself is immersed in a 14\,T solenoidal magnetic field. 
Semi-flexible RF-cables are connected to both cavity ports. The strongly coupled port, also named the readout port, is connected via a circulator (LNF-CIC4\_12A) to a 36\,dB cryo pre-amplifier (LNF-LNC4\_16B). 
The output of the preamplifier is connected to the readout electronics outside the cryostat. 
The second cavity port is strongly under-coupled and used to feed RF-signals into the cavity. 
This RF line is attenuated by 20\,dB close to the cavity in order to attenuate room-temperature noise from outside the cryostat. A third RF line is connected to the remaining port of the circulator, also attenuated by 20\,dB to inject signals into the readout port of the cavity.  Circulator and pre-amp are screwed onto a heat-exchanger to stabilize their temperature. The coax-cables are also connected to the heat-exchanger to allow thermalization of the cables in the lower part or the cryostat.

The temperature of the cavity and the heat-exchanger are monitored. The temperature is coarsely adjusted by controlling the LHe flow rate at a given pump-setting and stabilized by heating the bottom heat exchanger of the cryostat. The temperature is stable to better than 0.1\,K at 2\,K target temperature.

\begin{figure}[tb]
\centering
\begin{minipage}{8.2cm}
  \centering
    \includegraphics[width=0.99\linewidth]{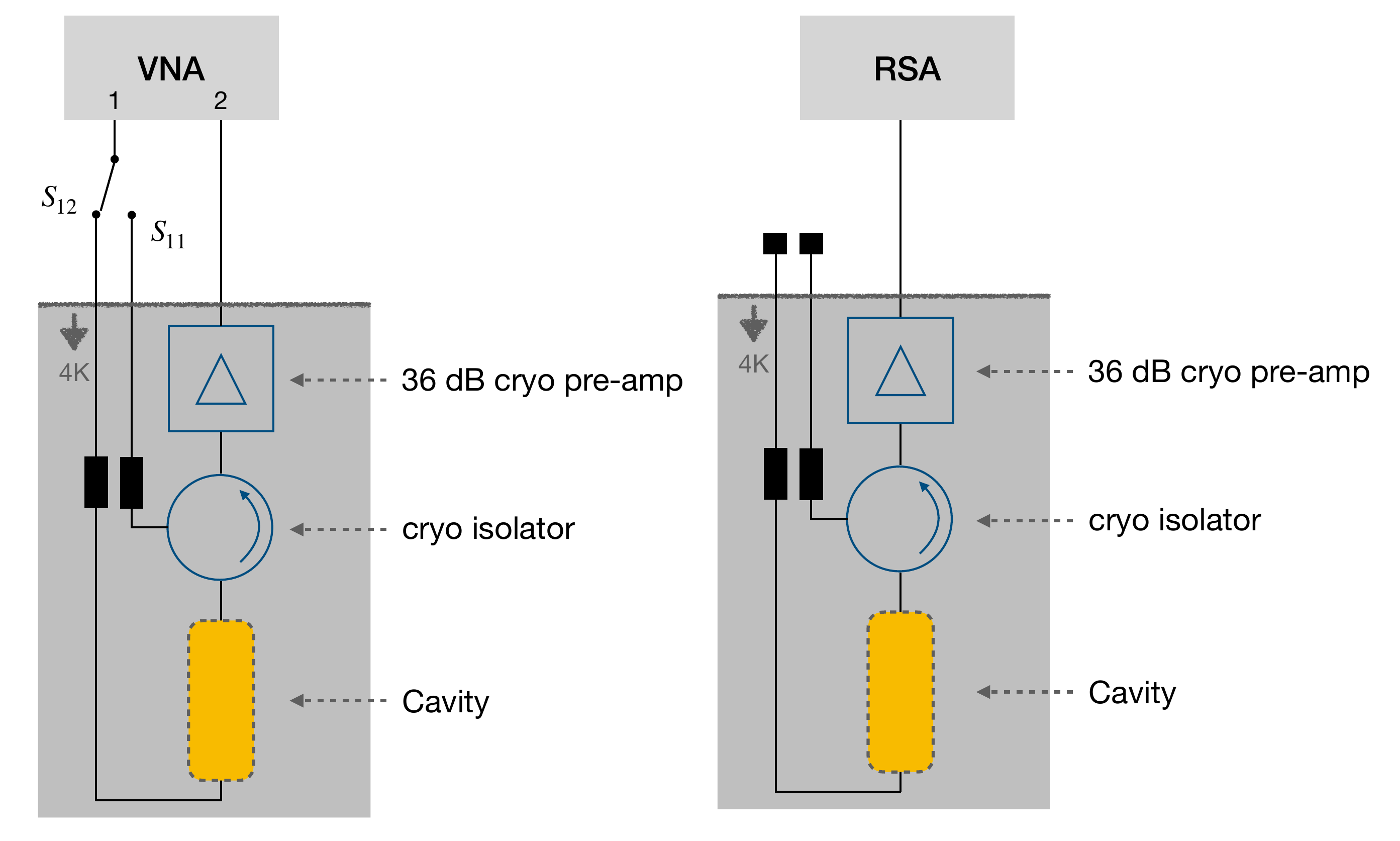}
    \caption{Schematic view of the readout path of the experiment. Left:  setup for S-Parameter measurement. Right: normal data taking mode.}
    \label{fig:DAQschematic}
\end{minipage}%
\hspace{0.2cm}
\begin{minipage}{8.2cm}
  \centering
    \includegraphics[width=0.99\linewidth]{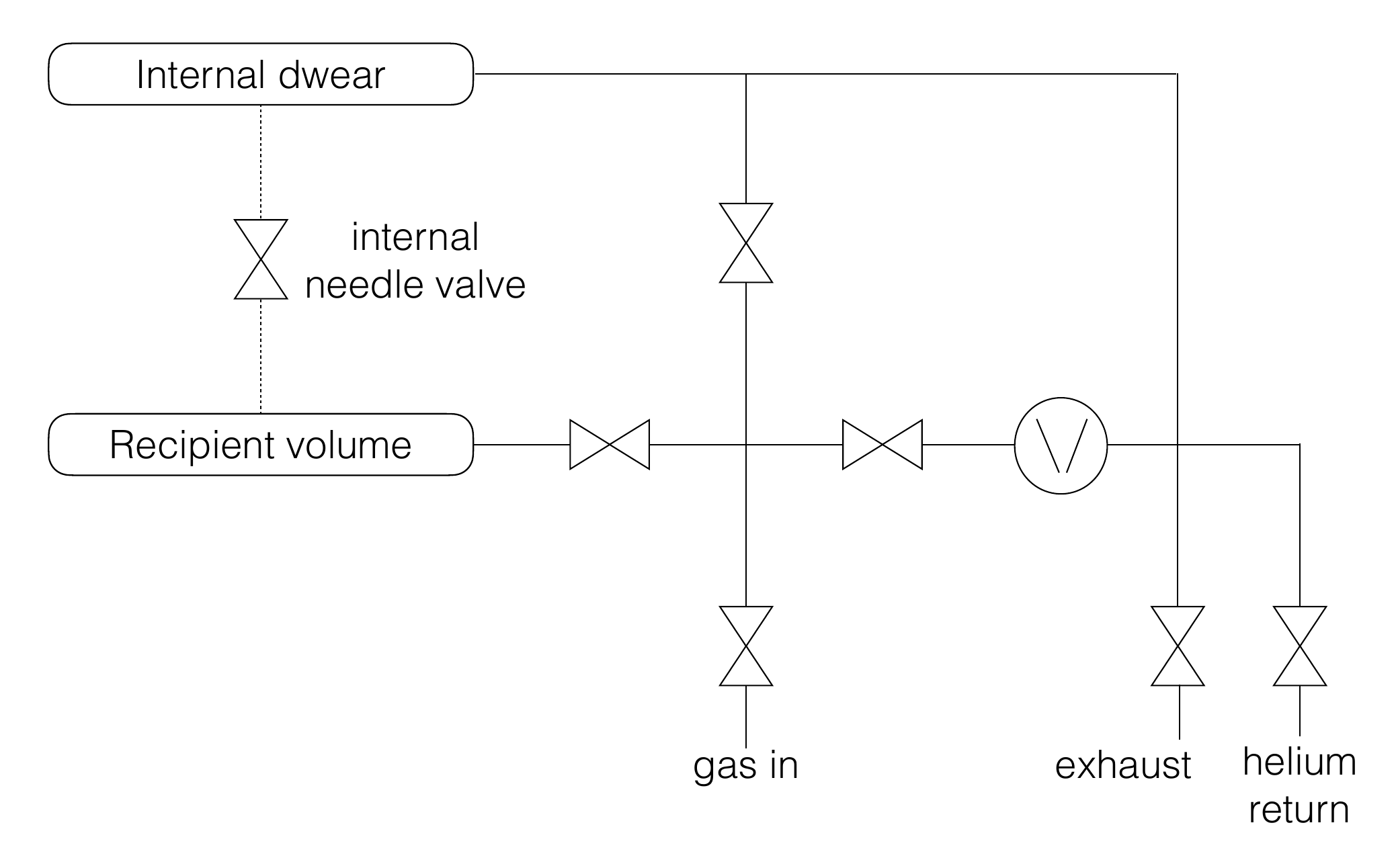}
    \caption{Schematic of the vacuum system used to control the pressure in the probe volume of the cryostat. This system is also used to evacuate the cryostat and flush it with helium gas before the cool-down. 
    \label{fig:vacuum-system}.}
\end{minipage}
\end{figure}

A picture of the cavity including the cryo readout electronics attached to the support structure is shown in Fig. \ref{fig:PictureInset}. Fig. \ref{fig:DAQschematic} shows a diagram of the readout electronics. 
For data taking the RF lines from the pre-amplifier is connected to a real-time spectrum analyzer
(Tektronix RSA 518A), which can digitize the RF signals in a bandwidth of 40\,MHz in real-time. 
The data are read-out to a PC via USB3, pre-processed and stored. For characterizing the cavity a network analyzer (Rhode\&Schwarz ZNB40) is connected to the cavity ports to perform a full set of S-parameter measurements.

\subsection{Cryostat and 2K operation}\label{sec:cryostat}
The LHe flow cryostat is operated at a pressure between 20\,mb to 100\,mb. To this end a vacuum system is needed, allowing to evacuate the whole system and flush it with helium gas as well to pump the cavity recipient with an adjustable rate. The vacuum system is shown in Fig. \ref{fig:vacuum-system}.
The cryostat features a liquid nitrogen shield to reduce LHe losses. 
Cool-down of the system from room temperature to 2\,K takes about 4h. After reaching 2\,K the LHe flow rate is minimized and the pump-rate adjusted to the reach the desired pressure. At that point the temperature is regulated via a heating element connected to a temperature controller. 
Changes of the pressure are realized by changing the LHe flow rate. To this end the LHe needle vale is actuated via a stepper motor allowing for automatic pressure control.
A turbo molecular pump is used to create an insulation vacuum with a remaining pressure of $< 10^{-5} mb$ before the liquid nitrogen shield is filled. The shield needs refilling every 6 hours in normal operation. 
For continuous operation the external LHe dwear remains connected to the internal dwear of the cryostat, allowing stable operation over several days. 
The external dwear can be replaced with minimal impact on the operation of system.

Pressure sensors are read out via an Arudino board while the temperature sensors are read by a CTC100 temperature controller which also controls the cryostat heater.
The environmental data is stored in an Influx database and visualized using Grafana displays for online-monitoring of the system. 



\subsection{Cavity properties}

A cavity with a resonance frequency of the TM$_{010}$ mode of 8.4~GHz was designed, based on experiences from the RADES group \cite{Golm:2021ooj}. The cavity is built out of two copper half-shells with dimensions 27~mm x 40~mm x 160~mm. The inner dimensions are 22.8~mm x 30~mm x 150~mm where all corners of the cavity are rounded with a radius of 9~mm. Expendable material is removed from the outside as much as possible to lower the heat capacity.
The final cavity weighs 643\,g.

\begin{figure} [ht]
    \centering
    \includegraphics[width=0.245\textwidth, angle=90]{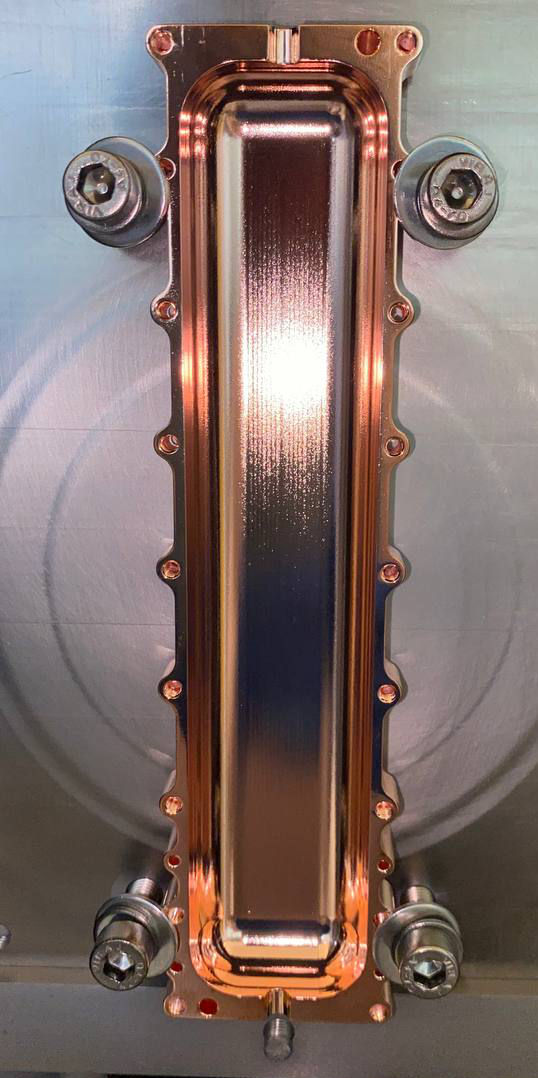}
    \hfill
    \includegraphics[width=0.245\textwidth, angle=90]{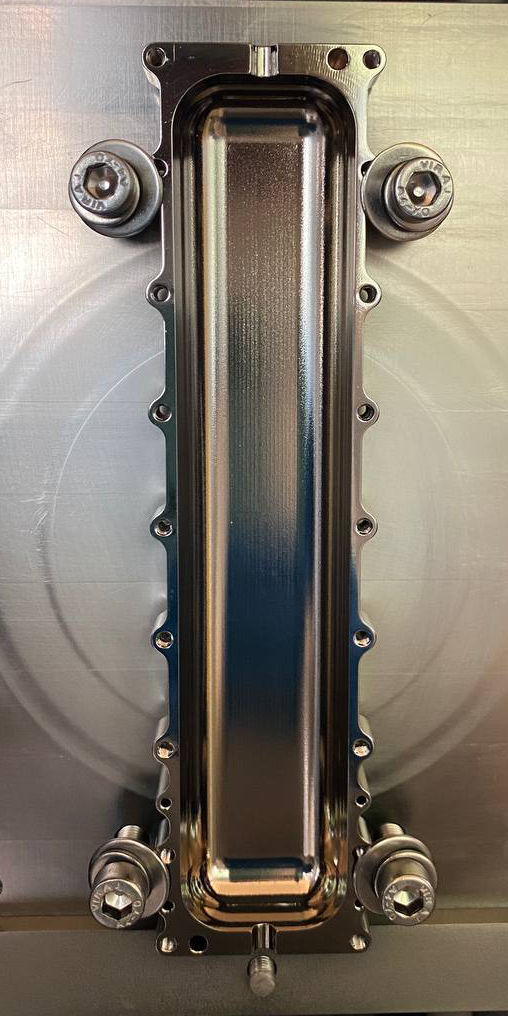}
    \caption{Images of the cavity under test. Left: the electropolished copper cavity. Right: after depositing the NbN superconductor.}
    \label{fig:cavity}
\end{figure}

Two cavities are built. One is coated with a $2\,\mu$m thick layer of NbN was deposited on the copper substrate \cite{Leith_2021, ubsi_2108} and used without external magnetic field, reaching a quality factor of $Q_0 = \Qnbn$.
The bare copper cavity with $Q_0 = \Qcu$ is used to search for axions in a 12\,T magnetic field. 
Images of the cavity half-shell before and after the coating are shown in Fig. 
\ref{fig:cavity}. Two antennas are attached to the cavity at opposite ends of its long-axis. One is strongly coupled to the TM$_{010}$ mode, the other weakly. 
The resonance frequency of the TM$_{010}$ mode is measured to be \fres at a temperature of 4~K.

\subsection{Heated load}\label{sec:heated-load}
To measure the system's noise temperature, a heated load is installed at the third port of the circulator. This setup enables the injection of thermal noise at a well-defined temperature into the cavity input. The heated load consists of a $50\,\Omega$ terminator mounted between the arms of a U-shaped copper bar with a thickness of 1.5\,mm. The copper bar is wrapped in alternating layers of Mylar tape and resistive wire. The total resistance of the wire is approximately  $10\,\Omega$, allowing for a maximum heating power of up to 10\,W. A DT670 temperature sensor has then been mounted directly on the copper bar to monitor the temperature, which is controlled via a CTC100 temperature controller. 

\section{Preliminary Results of the Prototype Experiment}

Dark photons are hypothetical gauge bosons associated with a hidden or "dark" sector, which may interact very weakly with Standard Model particles through kinetic mixing with ordinary photons. As such, they are compelling candidates for explaining dark matter and other phenomena beyond the Standard Model. In cavity-based experiments, dark photon searches exploit this mixing by looking for excess power deposited in a resonant electromagnetic cavity in the absence of any external signal. If dark photons make up a component of the local dark matter halo, they can convert into detectable photons inside a high-quality microwave cavity tuned to their mass (frequency), even without a magnetic field—unlike axion searches which require one. By scanning the resonance frequency of the cavity and monitoring for excess electromagnetic power, these experiments can probe previously unexplored regions of the dark photon parameter space, especially at microwave frequencies corresponding to $\mu$eV-scale masses. Hence our prototype setup could serve as an ideal experimental setup to search for dark-photon signatures without yet having the final magnet system installed. This search effort also allowed us to establish the full readout as well as analysis chain for the future \Supax experiment, as detailed in the following. 

\subsection{Data acquisition system and Data format}\label{sec:DAQ}
The RF signals from the cavity are amplified by 36\,db inside the cryostat. The amplified signals are fed into a real-time spectrum analyzer at room temperature. 
Typically a bandwidth of 10\,MHz around the cavity frequency is digitized. 
The IQ data is transferred to a standard PC via USB3 and processed in real-time. 

The time-domain IQ data is converted into frequency domain via a fast fourier transformation (FFT) where a Kaiser–Bessel window filter with a shape parameter of 2.5 is applied. The spectra resulting from a data stream of one second length are averaged on the fly and then stored as a ROOT \cite{Brun:1997pa} TGraph object. 
This leads to a low data rate to disk of 1.29\,MB/sec, while keeping the information on any drifts of the cavities RF properties over time. Data files of arbitrary length can be stored until disk space runs out.  In addition to the frequency domain data all settings of the RSA are stored in each output file as ROOT TTree object. To allow analysis of the time domain data the IQ-data stream can optionally be stored to disk. In this case the data rate at a readout bandwidth of 10\,MHz is about 100\,MB/sec. The DAQ software is written in C++ and relying on the Tektronix API for linux \cite{TektronixAPI} to communicate with the RSA.

\subsection{Frequency tuning via pressure changes }\label{sec:freq-tuning}
\begin{figure}[tb]
\centering
\begin{minipage}{7.2cm}
  \centering
    \includegraphics[width=0.99\linewidth]{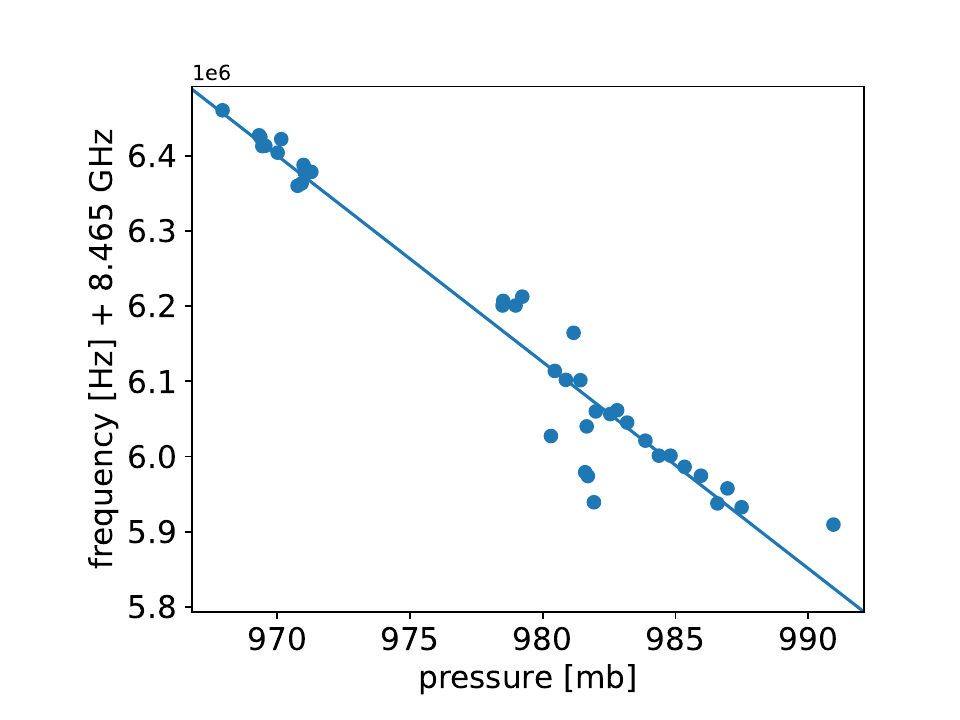}
    \caption{Dependence of the cavity's resonance frequency on the ambient He pressure. A linear fit yields a slope of \TuningSlopeNbN, while the cavity bandwidth is 28.5\,kHz.\vspace{0.05cm}}
    \label{fig:tuning_range}
\end{minipage}%
\hspace{0.2cm}
\begin{minipage}{9.2cm}
  \centering

    \includegraphics[width=0.99\linewidth]{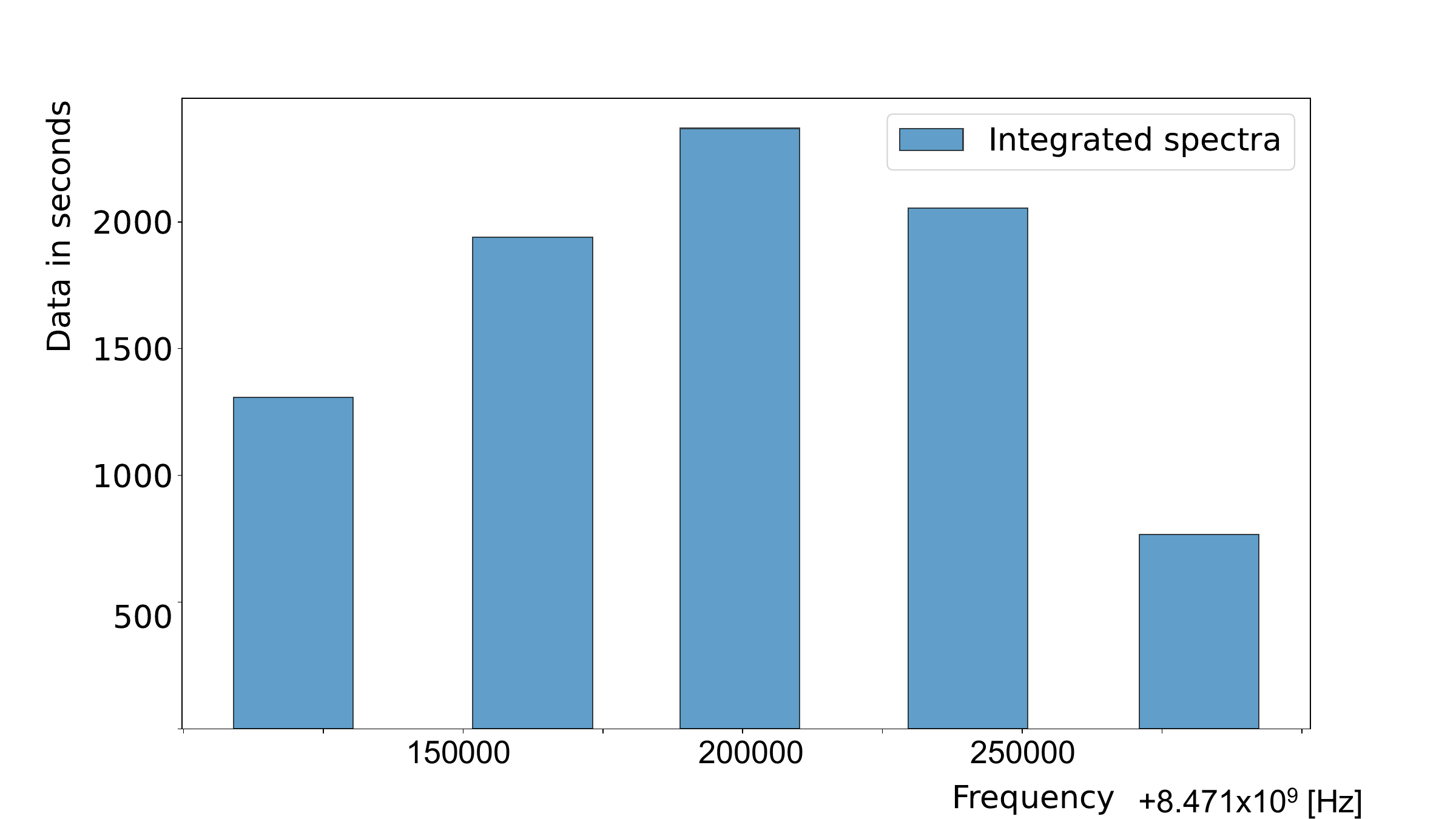}
    \caption{Shown is the amount of collected data in seconds per frequency interval. A minimum of 600 seconds is required per frequency bins to be considered in the analysis.}
    \label{fig:dataOverview}
\end{minipage}
\end{figure}

The cryostat allows operation at 2\,K in a significant range of pressures. Typically helium pressures in the recipient volume between 20\,mb and 100\,mb are used, while it can also be operated close to atmospheric pressure. 
The changing pressure in the cavity results in a changing He density in the cavity and hence a change in the relative permittivity of the cavity interior volume. This translate into a change of the resonance frequency of the cavity.
Figure \ref{fig:tuning_range} shows the relation of He pressure with resonance frequency yielding a linear dependence with a slope of \TuningSlopeNbN.
A change of 20\,mb in pressure allows for a tuning range of 550\,kHz. 
This is used to scan over a frequency range when searching for dark photons, as described in section \ref{sec:darkPhotonsAnalysis}.

Evidently, a significant tuning range is accessible via pressure changes at temperatures around 2\,K. 
This can be combined with a mechanical tuning of the resonance frequency, allowing for a wide tuning range. Given the precision of the pressure tuning the mechanical tuning part can be in coarse steps, significantly simplifying the needed mechanical setup.


%
%

\subsection{Analysis and Results}\label{sec:darkPhotonsAnalysis}

In October 2024 a measurement was performed utilizing a superconducting cavity with a quality factor of $Q=3\cdot 10^5$. The characterization of the cavity is shown in detail in \cite{Schmieden:2024wqp}. By changing the pressure a frequency range  of \scanRange around the center frequency of \fres was scanned. The integration time was at least 750 sec per frequency interval. 
 Frequency intervals are defined to have a width corresponding to the cavity bandwidth of 
 $BW = \frac{f_0}{Q} = 28.5\,kHz$. Fig. \ref{fig:dataOverview} shows the collected integration time per frequency interval.
 Only measurements at times where pressure and temperature of the setup were stable are considered. 
As described in section \ref{sec:DAQ} the time domain data is converted into frequency space in real-time and averaged in intervals of one second. 
The resulting spectra are the basis for the data analysis which closely follows the analysis procedure outlined in \cite{Brubaker:2017rna}.

The data analysis is carried out in four main steps, each designed to progressively refine the signal and extract potential axion-induced features. First, the raw data is integrated to improve the signal-to-noise ratio. In the second step, the intrinsic cavity response and artifacts introduced by the readout electronics are carefully removed to isolate physical signals from instrumental effects. Next, adjacent frequency bins are combined to optimize sensitivity across the scanned spectrum. Finally, statistical methods are applied to set exclusion limits on the axion-photon coupling strength. Each of these steps is explained in detail in the following.

\begin{figure}[ht]
    \centering
    \includegraphics[width=0.48\linewidth]{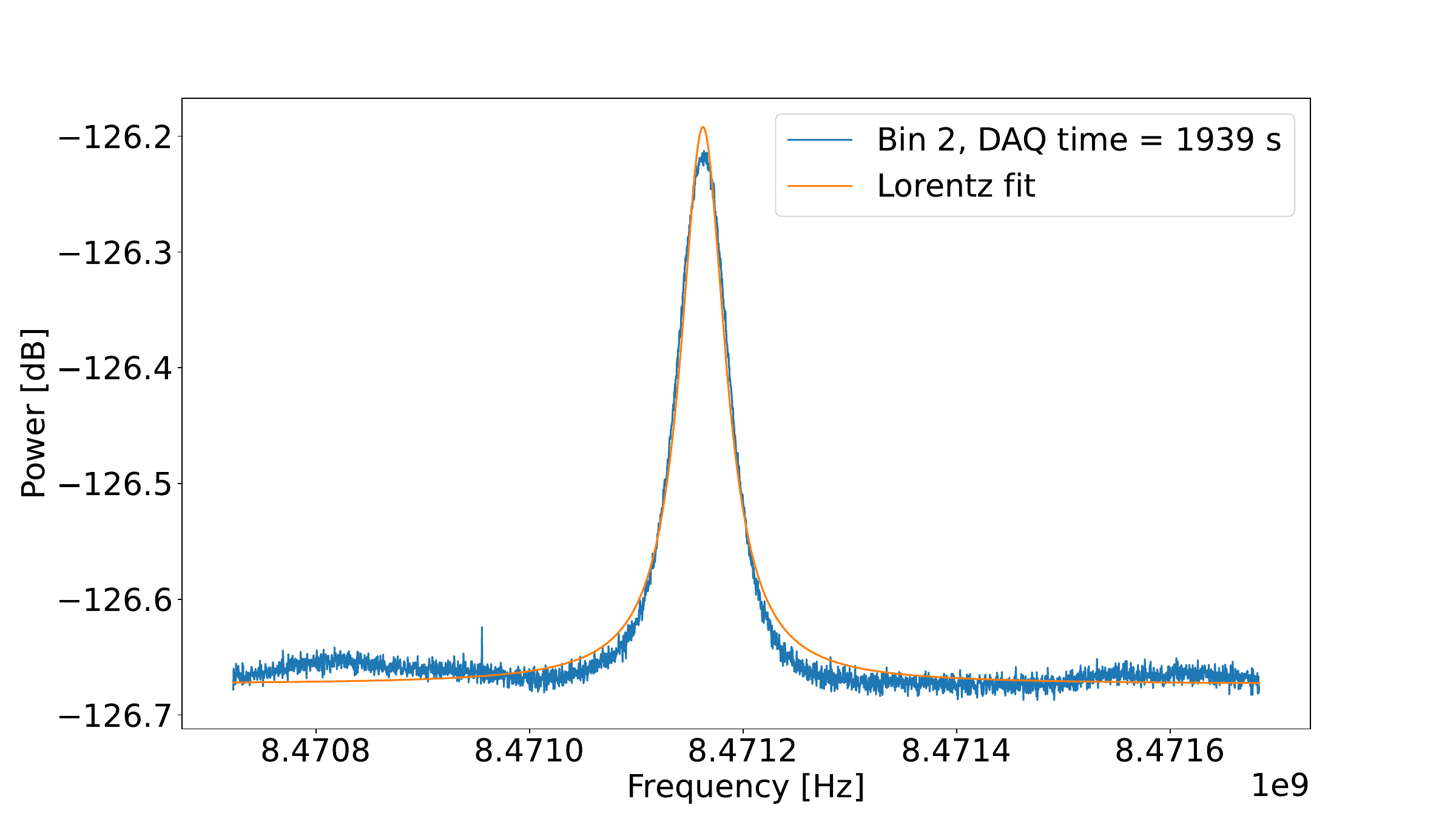}
    \includegraphics[width=0.48\linewidth]{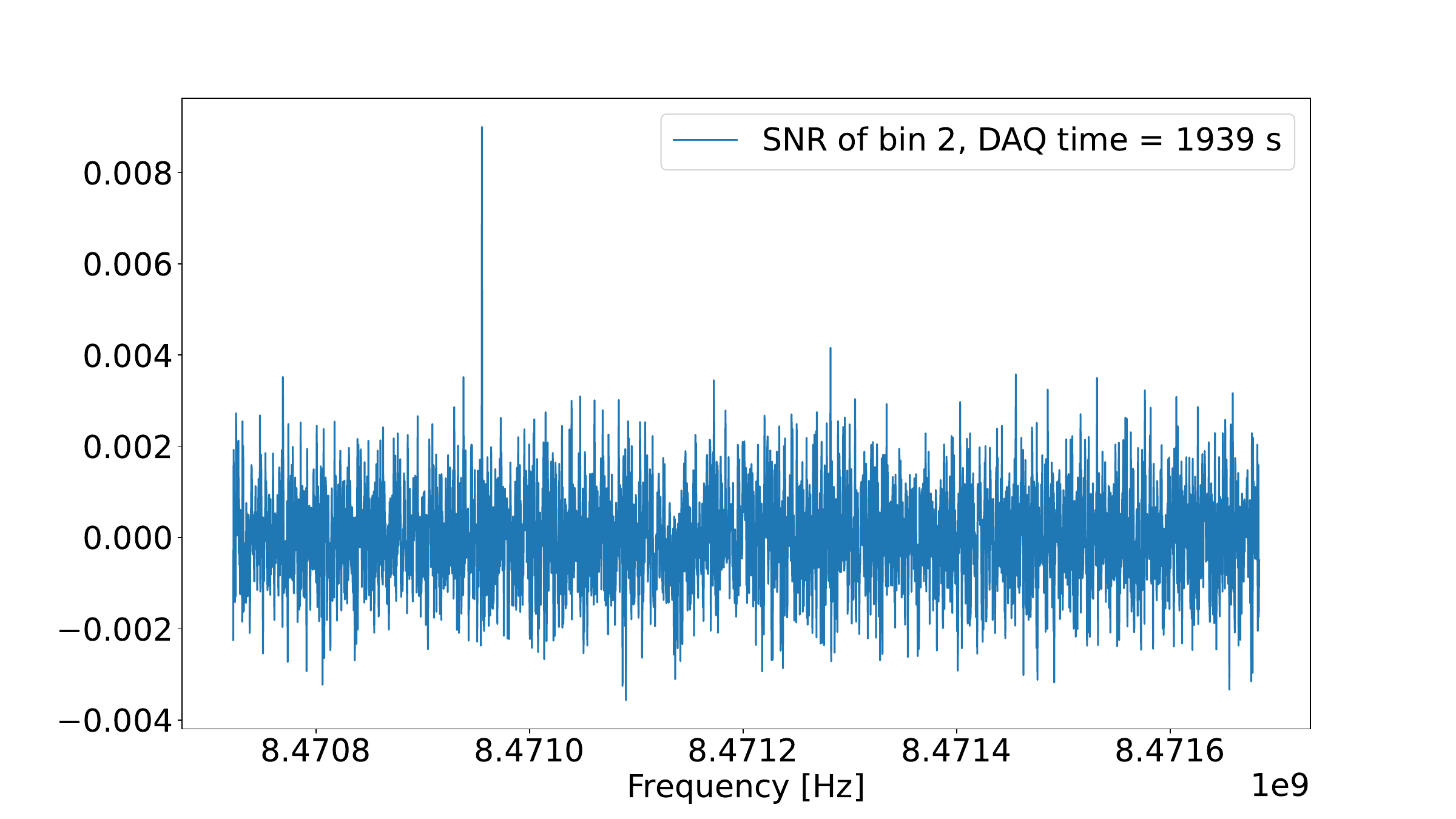}
    \includegraphics[width=0.48\linewidth]{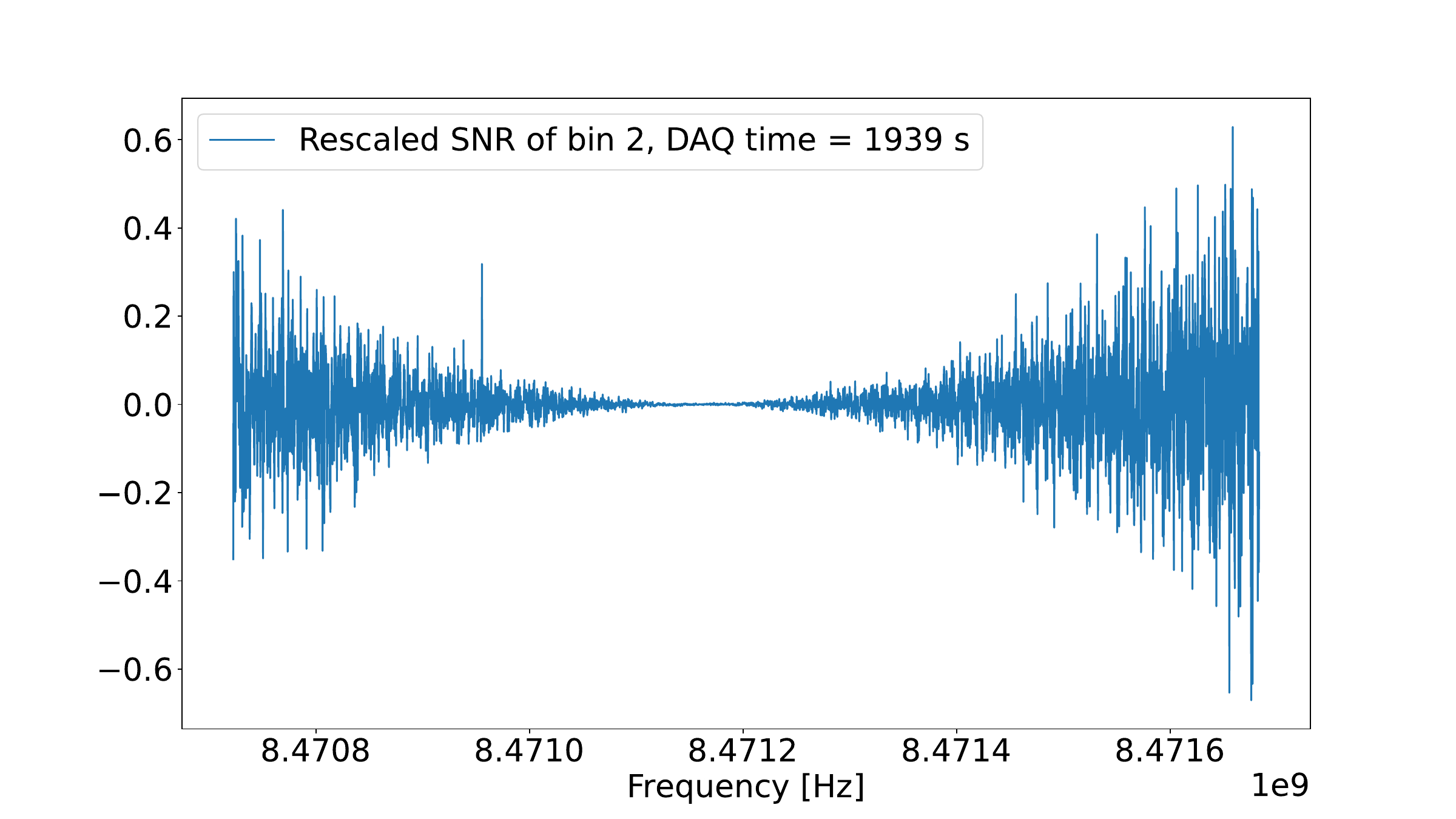}
    \includegraphics[width=0.48\linewidth]{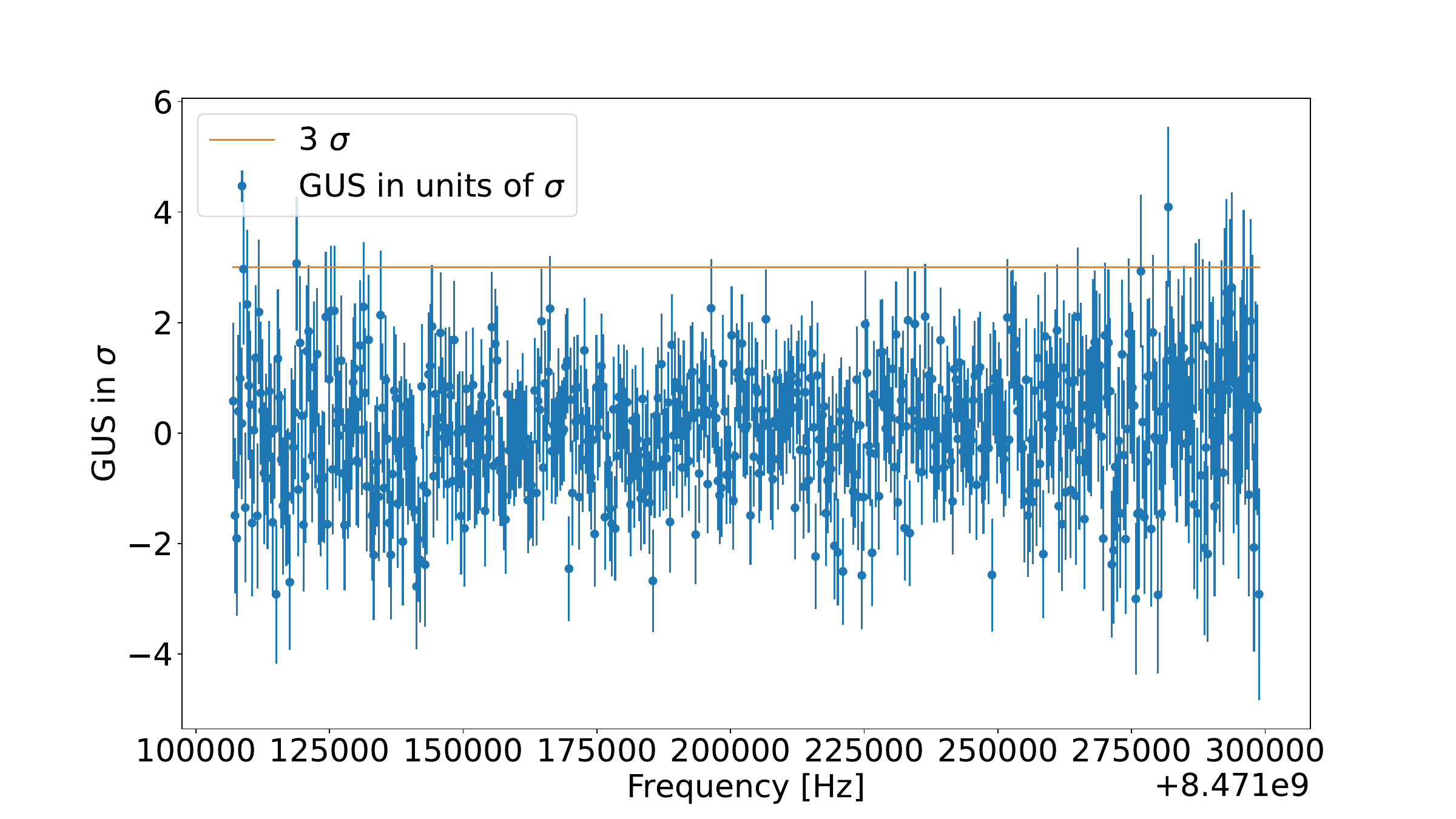}
    \caption{Plots showing the different analysis steps exemplary for frequency bin 2. Upper left: Measured power spectrum integrated over 30 min. Upper right: Power spectrum divided by the SG-Filter result and shifted by -1, called SNR. Lower left: Rescaled power SNR spectrum. Lower right: central frequency rage of the grand unified spectrum, where all frequency bins are combined. }
    \label{fig:DataProcessing}
\end{figure}

For each recorded spectrum of one second length the resonance frequency is estimated by fitting the resonance curve of the cavity which is visible as the readout port of the cavity is not exactly critically coupled. 
Spectra where the resonance frequency is within a narrow range of 28.5\,kHz, which corresponds to the cavity band width, are averaged yielding a spectrum as shown in Fig. \ref{fig:DataProcessing} (a) with significantly reduce noise. 
This is called the \emph{raw} spectrum.
In order to remove the resonance structure of the cavity as well as any remaining artifact of the DAQ system the raw spectrum is processed with a Savitzky–Golay (SG) filter \cite{Savitzky:2002vxy} and divided by the result and shifted by -1 to be centered around 0. 
The resulting dimensionless spectrum is labeled \emph{SNR} and shown in Fig. \ref{fig:DataProcessing} (b). 
Each processed spectrum is multiplied by the average noise power per frequency bin and divided by the dark photon conversion power profile to obtain a set of \emph{rescaled spectra} shown in Fig. \ref{fig:DataProcessing} (c).
Finally, the rescaled spectra from all frequency bins are combined by taking an optimally weighted sum across the whole frequency range. The result is called the \emph{grand unified spectrum} shown in \ref{fig:DataProcessing} (d).

\begin{figure}[tb]
\centering
\begin{minipage}{8.2cm}
  \centering
    \includegraphics[width=0.99\linewidth]{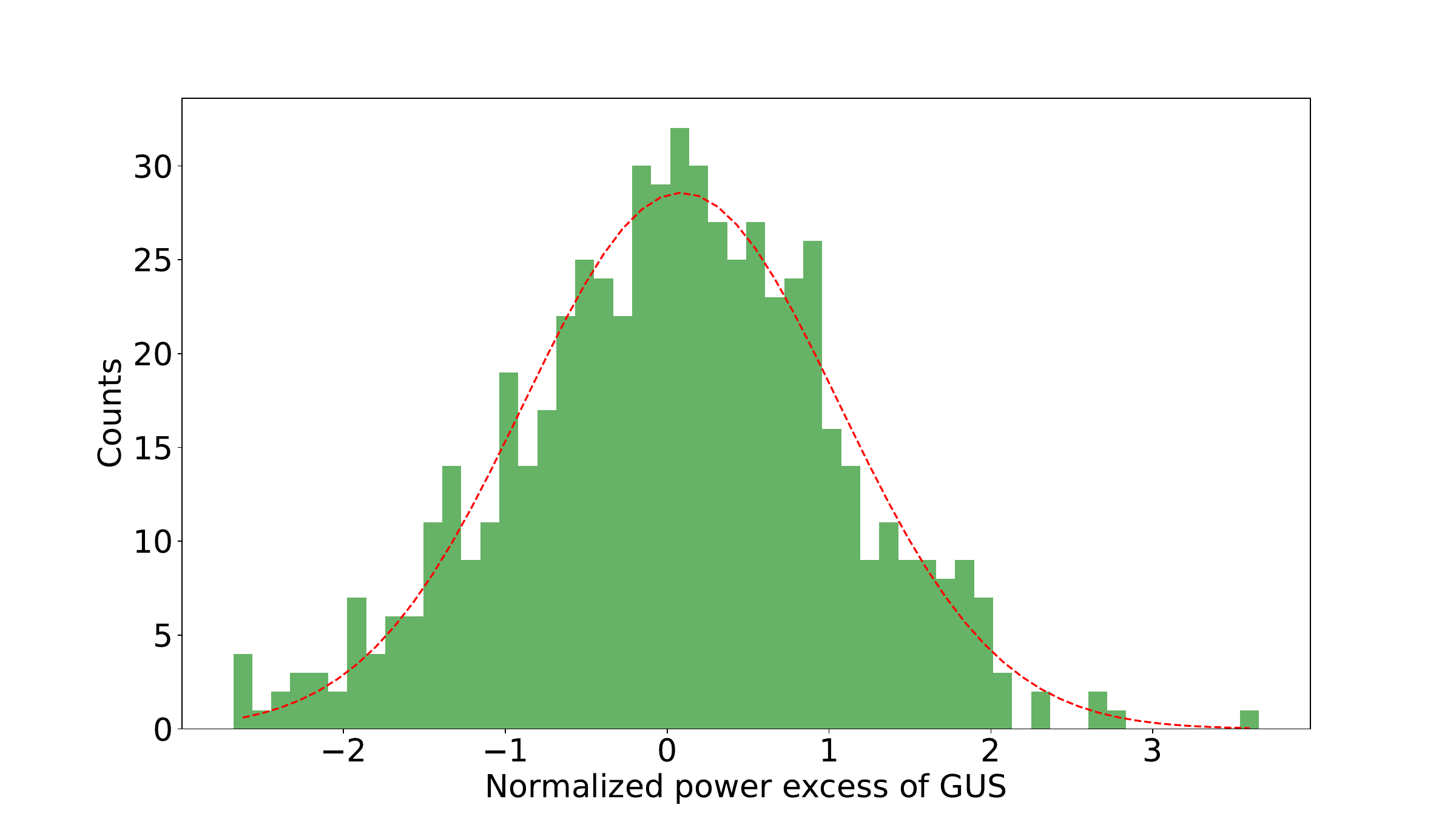}
    \caption{Distribution of the normalized SNR entries in the grand unified spectrum. As can be seen the distribution follows nicely a gaussian distribution, as expected.\vspace{0.35cm}}
    \label{fig:normalizedSpectrum}
\end{minipage}%
\hspace{0.2cm}
\begin{minipage}{8.2cm}
  \centering
    \includegraphics[width=0.85\linewidth]{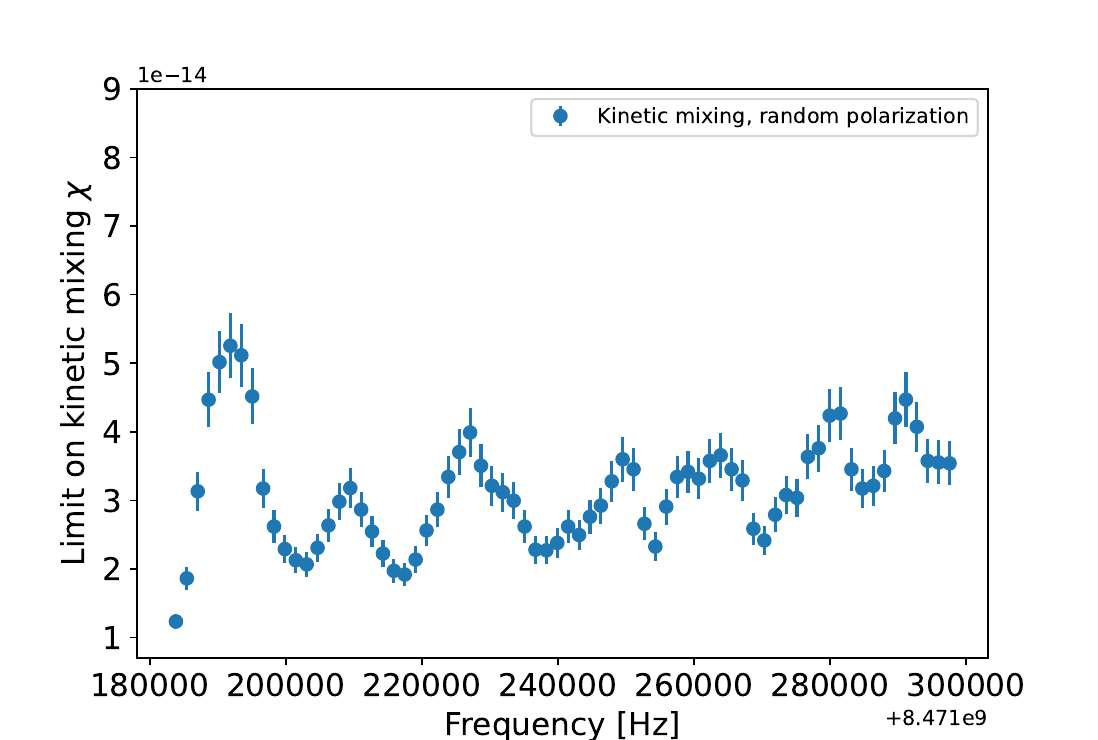}
    \caption{Measured upper limit on the kinetic mixing parameter $\chi$ in dependence on the resonance frequency of the cavity assuming a random polarization of the dark photon. No significant excess is observed}
    \label{fig:limits_finebinned}
\end{minipage}
\end{figure}

\begin{figure}[ht]
    \centering
    \includegraphics[width=0.85\linewidth]{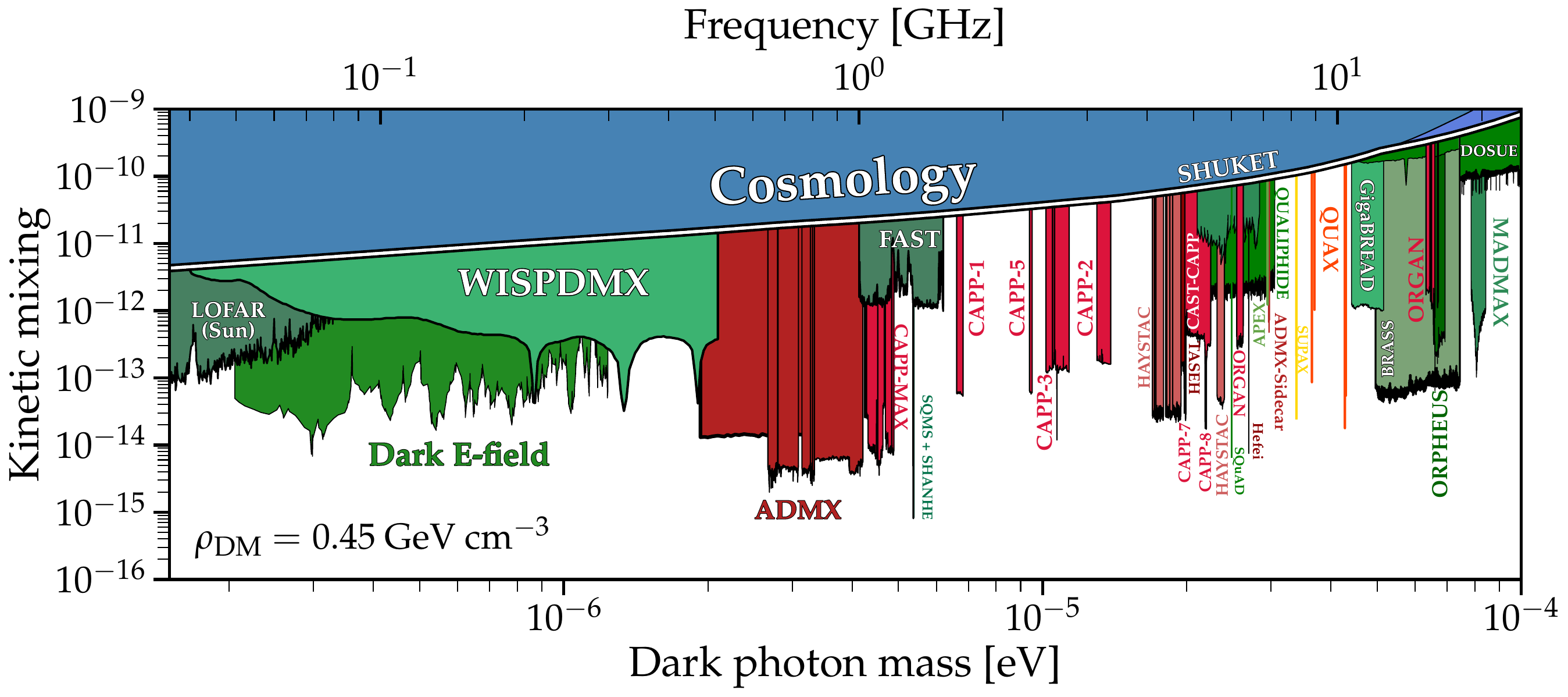}
    \caption{Comparison of the measured limits in comparison to other results taken from \cite{AxionLimits}.}
    \label{fig:UpperLimits}
\end{figure}

In the absence of any signal, the normalized spectrum as well as the grand unified spectrum exhibit a gaussian distribution of the normalized power entries, as shown in Fig. \ref{fig:normalizedSpectrum}.
The confidence level at which dark photon candidates can be excluded at a certain value of the kinetic mixing $\chi$ is set to $c_1 = 95\%$. 
A threshold in the SNR spectrum has to be defined above which an excess is considered a potential signal. Each excess has then to be scanned again to exclude a statistical fluctuation. 
For this initial measurement no re-scans were performed. Hence the threshold $\Theta$ for performing a re-scan is set to the largest observed fluctuation at $4.1\sigma$. 
The chosen confidence level and the re-scan threshold define the sensitivity to the dark-photon kinetic mixing parameter via 
\begin{equation}
    |\chi^{min}|_l = G_l|\chi^{\text{0}}|, \qquad G_l = \sqrt{ R_T / \tilde{R}_l^g},
\end{equation}
where $\tilde{R}_l^g$ is the SNR for an dark-photon signal with $|\chi| = |\chi^{\text{0}}|$ and $_l$ denotes the frequency bin.
$R_T$ is the target SNR defined as $R_T = \Theta + \Phi^{-1}(c_1) = 5.7$. 
The resulting limits on the kinetic mixing parameter are presented in Fig. \ref{fig:limits_finebinned} and range between $\chi < 1\cdot 10^{-14}$ to  $\chi < 5\cdot 10^{-14}$ in a frequency band of 150\,kHz around $f_0 = 8.4711\,$GHz.
The comparison to other experiments is presented in Fig. \ref{fig:UpperLimits}.


\section{Conclusion}

The SUPerconduction AXion search experiment represents a significant advancement in the search for axion-like particles (ALPs) and dark photons, two well-motivated candidates for physics beyond the Standard Model. By combining a high-field 12\,T superconducting solenoid with a novel cavity design featuring movable central dielectric elements and superconducting surfaces, \Supax is capable of scanning axion masses in the range of \AxionMassMin to \AxionMassMax, corresponding to resonance frequencies between 2\,GHz and 7.2\,GHz. A key innovation of the experiment is the multi-cavity structure within a single copper body, where tuning is achieved through an off-center dielectric rod actuated by piezo-actuators. This allows for simultaneous, high-precision frequency tuning of both cavities. Complementary fine-tuning is realized through controlled helium gas pressure variations, enabling frequency adjustments on the order of tens of kHz. In addition to axion searches, the \Supax setup is inherently sensitive to dark photon dark matter through its high-quality cavity design, without the need for an external magnetic field. The experiment's prototype phase, carried out at the University of Mainz, validated the cryogenic and RF systems, and already places new constraints on dark photons in a previously unexplored parameter space. With its modular, scalable design and strong sensitivity projections, \Supax lays the groundwork for a competitive and flexible platform for next-generation searches for ALPs, dark photons, and potentially other hidden sector particles.

\section*{Acknowledgement}
The authors thank the research group of Dmitri Budker at the University of Mainz, in particular Malavika Unni, Hendrik Bekker and Arne Wickenbrock for their support during our test measurements. The work of Tim Schneemann was funded by the DFG graduate school "modern detectors".

\bibliographystyle{apsrev4-1} 
\bibliography{./Bibliography}
	
\end{document}